\documentclass[]{ptptex}
\usepackage{wrapft}
\usepackage{graphicx}

\newcommand{\ket}[1]{| {#1} \rangle}     
\newcommand{\rket}[1]{| {#1} )}     
\newcommand{\rdket}[1]{|\!| {#1} )}     

\newcommand{\wtilde}[1]{\widetilde{#1}} 

\def\beq{\begin{eqnarray}}
\def\eeq{\end{eqnarray}}
\def\bsub{\begin{subequations}}
\def\esub{\end{subequations}}
\def\b{\begin{equation}}

\title{
Color-Singlet Three-Quark States in the $su(4)$-Algebraic Many-Quark Model
}
\subtitle{
An Example of the $su(4)\otimes su(4)$-Model
}    

\author{
Yasuhiko {\sc Tsue},$^{1}$ 
Constan\c{c}a {\sc Provid\^encia},$^{2}$ 
Jo\~ao da {\sc Provid\^encia}$^{2}$ and 
Masatoshi {\sc Yamamura}$^{3}$  
}

\inst{
$^{1}$Physics Division, Faculty of Science, Kochi University, Kochi 780-8520, 
Japan\\
$^{2}$Departamento de F\'{i}sica, Universidade de Coimbra, 3004-516 Coimbra, 
Portugal\\
$^{3}$Department of Pure and Applied Physics, 
Faculty of Engineering Science,\\
Kansai University, Suita 564-8680, Japan
\\

}

\recdate{
\today
}

\abst{
As announced in last paper by the present authors, the color-singlet three-quark states 
are investigated for the case of the spherical $j$-$j$ coupling and 
$L$-$S$ coupling shell models. 
The latter case automatically leads to the $su(4)\otimes su(4)$-model. 
For a certain type of Hamiltonian, the states of individual ``nucleons" are characterized.
}

\begin{document}

\maketitle


The $su(4)$-algebraic quark model, which is an example of a many-fermion model, 
may be quite attractive. 
With the aid of this model, we can learn various aspects of a many-quark system, 
for example, the quark-triplet formation and the pairing correlation. 
The first aspect is related to describing the structure of the ``nucleon" and the second one  
to describing color-superconductivity. 
In our last paper,\cite{0} which will be referred to as (I), 
we presented a possible systematic treatment of this model. 
In \S 6 of (I), we derived two forms which result in the ``nucleon". 
Both are based on the $j$-$j$ coupling and the $L$-$S$ coupling shell model 
and the explicit forms of the former and the latter are shown in the relations (I$\cdot$6$\cdot$3) 
and (I$\cdot$6$\cdot$6), respectively. 
The form (I$\cdot$6$\cdot$3) is of the same form as that presented by Petry et al.\cite{2} 
and the form (I$\cdot$6$\cdot$6) is derived under our idea. 
Main aim of this note is to give our own interpretation of the form (I$\cdot$6$\cdot$3) 
in connection with the form (I$\cdot$6$\cdot$6). 
We will not present a reinterpretation of the notation appearing in (I).

Let us start by rewriting the first quark-triplet operator in (I$\cdot$6$\cdot$6) in spherical 
tensor representation. 
For this aim, we use the quantum number $l_s$ explicitly, but omit $1/2$ for the spin and the isospin. 
The first quark-triplet operator in (I$\cdot$6$\cdot$6) can be expressed as follows: 
\beq
& &{\wtilde B}_{l_s\mu\sigma\tau}^*=\sum_{k=1}^3{\wtilde S}^k(l_s){\tilde c}_{kl_s\mu\sigma\tau}^* \ , 
\label{1}\\
& &{\wtilde S}^1(l_s)=\sum_{\mu\sigma\tau}(-)^{l_s-\mu}(-)^{\frac{1}{2}-\sigma}(-)^{\frac{1}{2}-\tau}
{\tilde c}_{2l_s\mu\sigma\tau}^*{\tilde c}_{3l_s{\tilde \mu}{\tilde \sigma}{\tilde \tau}}^*\ . 
\label{2}
\eeq
The cases $k=2$ and 3 are obtained by cyclic permutation of the color 
$i=1$, 2 and 3. 
By considering the addition of the orbital and the spin angular momentum, 
we can define ${\tilde c}_{kl_sj_{\pm}m\tau}^*$ in the form 
\beq\label{3}
{\tilde c}_{kl_sj_{\pm}m\tau}^*
=\sum_{\mu\sigma}\langle l_s\mu 1/2 \sigma\ket{j_{\pm}m}{\tilde c}_{kl_s\mu\sigma\tau}^* \ . 
\eeq
Here, $\langle l_s\mu 1/2 \sigma\ket{j_{\pm}m}$ denotes the Clebsch-Gordan coefficient and $j_{\pm}$ are given as 
\beq\label{4}
j_+=l_s+1/2\ , \qquad j_-=l_s-1/2\ . \quad (l_s\geq 1)
\eeq
Since ${\wtilde S}^k(l_s)$ is scalar, ${\wtilde B}_{l_sj_{\pm}m\tau}^*$ can be defined as 
\beq\label{5}
{\wtilde B}_{l_sj_{\pm}m\tau}^*=\sum_{\mu\sigma}\langle l_s \mu 1/2 \sigma \ket{j_{\pm}m}{\wtilde B}_{l_s \mu\sigma\tau}^*\ .
\eeq
Then, we can prove the following relation: 
\beq\label{6}
{\wtilde B}_{l_sj_{\pm}m\tau}^*=\sum_{k=1}^3
\left({\wtilde S}^k(l_sj_+)+{\wtilde S}^k(l_sj_-)\right){\tilde c}_{kl_sj_{\pm}m\tau}^*\ . 
\eeq
Here, ${\wtilde S}^1(l_sj_{\pm})$ is defined as 
\beq\label{7}
{\wtilde S}^1(l_sj_{\pm})=\sum_{m\tau}(-)^{j_{\pm}-m}(-)^{\frac{1}{2}-\tau}{\tilde c}_{2l_sj_{\pm}m\tau}^*{\tilde c}_{3l_sj_{\pm}{\tilde m}{\tilde \tau}}^*\ . 
\eeq
The cases $k=2$ and 3 are obtained by cyclic permutation of the color 
$i=1$, 2 and 3. 
Since we are treating two single-particle levels specified by $(l_s,j_+)$ and $(l_s, j_-)$, the 
present $su(4)$-algebra is an $su(4)\otimes su(4)$-algebra: 
\bsub\label{8}
\beq
& &{\wtilde S}^i(l_s)={\wtilde S}^i(l_sj_+)+{\wtilde S}^i(l_sj_-)\ , 
\label{8a}\\
& &{\wtilde S}_i^j(l_s)={\wtilde S}_i^j(l_sj_+)+{\wtilde S}_i^j(l_sj_-)\ . 
\label{8b}
\eeq
\esub
The levels specified by $(l_s, j_{\pm})$ will be called the spin-orbit partner levels.

Next, we consider the first quark-triplet operator in (I$\cdot$6$\cdot$3). 
In the shell model, the quantum number $j_{\pm}$ is obtained by adding the orbital and the 
spin angular momentum, $l_s$ and $1/2$. 
In the case of the $L$-$S$ coupling scheme, 
the spin-orbit partner levels 
are expected to be close to each other and 
they are strongly correlated. 
The relation (\ref{6}) tells that ${\wtilde B}_{l_sj_{\pm}m\tau}^*$ depends on the 
two forms ${\wtilde S}^k(l_sj_+)$ and ${\wtilde S}^k(l_sj_-)$ in an equal weight. 
But, in the case of the $j$-$j$ coupling scheme, the two levels are expected to be 
separated and they are weakly correlated. 
Then, we can treat ${\wtilde B}_{l_sj_+m\tau}^*$ independently of 
${\wtilde S}^k(l_sj_-)$ and vice versa. 
The above consideration allows to rewrite the first quark-triplet operator in 
(I$\cdot$6$\cdot$3) in the following form: 
\beq\label{9}
{\wtilde B}_{l_sj_{\pm}m\tau}^*=\sum_{k=1}^3{\wtilde S}^k(l_sj_{\pm}){\tilde c}_{kl_sj_{\pm}m\tau}^*\ . 
\eeq
As was already mentioned, 
the main aim of this note is to discuss the form (\ref{9}) in connection with the form (\ref{6}).

Let us define the following four operators:
\bsub\label{10}
\beq
& &{\wtilde B}_{l_sj_+m\tau}^*(+)=\sum_k{\wtilde S}^k(l_sj_+){\tilde c}_{kl_sj_+m\tau}^* \ , 
\label{10a}\\
& &{\wtilde B}_{l_sj_+m\tau}^*(-)=\sum_k{\wtilde S}^k(l_sj_-){\tilde c}_{kl_sj_+m\tau}^* \ , 
\label{10b}
\eeq
\esub
\bsub\label{11}
\beq
& &{\wtilde B}_{l_sj_-m\tau}^*(-)=\sum_k{\wtilde S}^k(l_sj_-){\tilde c}_{kl_sj_-m\tau}^* \ , 
\label{11a}\\
& &{\wtilde B}_{l_sj_-m\tau}^*(+)=\sum_k{\wtilde S}^k(l_sj_+){\tilde c}_{kl_sj_-m\tau}^* \ . 
\label{11b}
\eeq
\esub
The above four operators commute with ${\wtilde S}_i^j(l_s)$ ($i\neq j$) defined 
by the relation (\ref{8b}) and, thus, they are color-singlet. 
The relations (\ref{6}) and (\ref{9}) can be expressed in the form 
\bsub\label{12}
\beq
& &{\wtilde B}_{l_sj_+m\tau}^{*(A)}={\wtilde B}_{l_sj_+m\tau}^*(+)+{\wtilde B}_{l_sj_+m\tau}^*(-) \ , 
\label{12a}\\
& &{\wtilde B}_{l_sj_-m\tau}^{*(A)}={\wtilde B}_{l_sj_-m\tau}^*(-)+{\wtilde B}_{l_sj_-m\tau}^*(+) \ , 
\label{12b}
\eeq
\esub
\vspace{-0.6cm}
\bsub\label{13}
\beq
& &{\wtilde B}_{l_sj_+m\tau}^{*(B)}={\wtilde B}_{l_sj_+m\tau}^*(+)\ , 
\qquad\qquad\qquad
\label{13a}\\
& &{\wtilde B}_{l_sj_-m\tau}^{*(B)}={\wtilde B}_{l_sj_-m\tau}^*(-)\ . 
\qquad\qquad\quad
\label{13b}
\eeq
\esub
In order to discriminate between both forms, we used the superscripts $(A)$ and $(B)$. 
It may be interesting to search under which condition the forms (\ref{12}) and (\ref{13}) are 
realized for a given Hamiltonian. 

\begin{figure}[b]
\begin{center}
\includegraphics[height=2cm]{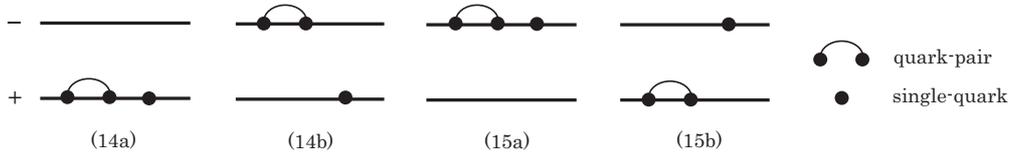}
\caption{Configurations of three quarks in one ``nucleon" in the states (\ref{14}) and (\ref{15}) 
are shown illustratively.}
\label{fig:1}
\end{center}
\end{figure}

We present an example illustrative of the idea how to treat the simplest case of the 
quark-triplet formation as a color-singlet, namely, one ``nucleon" problem. 
First, we introduce the following orthonormalized color-singlet states consisting of 
three quarks: 
\bsub\label{14}
\beq
& &\rket{l_sj_+m\tau;+}=(\sqrt{6(l_s+2)})^{-1}{\wtilde B}_{l_sj_+m\tau}^*(+)\rket{0}\ (=\rket{j_+;+}) \ , 
\label{14a}\\
& &\rket{l_sj_+m\tau;-}=(\sqrt{6l_s})^{-1}{\wtilde B}_{l_sj_+m\tau}^*(-)\rket{0}\ (=\rket{j_+;-}) \ , 
\label{14b}
\eeq
\esub
\vspace{-0.6cm}
\bsub\label{15}
\beq
& &\rket{l_sj_-m\tau;-}=(\sqrt{6(l_s+1)})^{-1}{\wtilde B}_{l_sj_-m\tau}^*(-)\rket{0}\ (=\rket{j_-;-}) \ , 
\label{15a}\\
& &\rket{l_sj_-m\tau;+}=(\sqrt{6(l_s+1)})^{-1}{\wtilde B}_{l_sj_-m\tau}^*(+)\rket{0}\ (=\rket{j_-;+}) \ . 
\label{15b}
\eeq
\esub
The above four states can be shown pictorially in Fig.{\ref{fig:1}.
In the frame of the above orthonormalized states, we investigate 
the following Hamiltonian: 
\beq\label{16}
{\wtilde {\cal H}}={\wtilde H}_0+{\wtilde H}_1 \ , 
\qquad\qquad\qquad\qquad\qquad\qquad
\eeq
\vspace{-0.6cm}
\bsub\label{17}
\beq
& &{\wtilde H}_0=g_0(l_s+1){\wtilde N}_{l_0}(-)-g_0l_s{\wtilde N}_{l_0}(+)\ , \qquad (g_0 \geq 0) 
\label{17a}\\
& &{\wtilde H}_1=-g_1\sum_{i=1}^3{\wtilde S}^i(l_s){\wtilde S}_i(l_s)\ . \qquad\qquad\qquad\quad (g_1 \geq 0)
\label{17b}
\eeq
\esub
Here, ${\wtilde N}_{l_0}(-)$ and ${\wtilde N}_{l_0}(+)$ denote quark-number operators of the states 
specified by $(-)$ and $(+)$, respectively,  and 
$g_0$ and $g_1$ denote parameters. 
The first part ${\wtilde H}_0$ comes from the spin-orbit interaction for the quark, 
$-2g_0{\mib l}\cdot{\mib s}$ and it may be natural for our two-level model to adopt this interaction. 
The second part ${\wtilde H}_1$ characterizes the quark-triplet formation. 
We will search the eigenstates of the Hamiltonian (\ref{16}). 
As is clear from the states (\ref{14a}) and (\ref{14b}), the Hamiltonian (\ref{16}) has 
two eigenstates for a given $j_+$ and the same happens in the case $j_-$. 
We pay an attention to the state with the lower energy 
and require that this state is one ``nucleon" state. 
The operation of ${\wtilde {\cal H}}$ on the above four states gives us the following results: 
\bsub\label{18}
\beq
& &{\wtilde {\cal H}}\rket{j_+;+}=-\left[
3g_0l_s+2g_1(l_s+2)\right]\rket{j_+;+}-2g_1\sqrt{l_s(l_s+2)}\rket{j_+;-} \ , 
\label{18a}\\
& &{\wtilde {\cal H}}\rket{j_+;-}=-2g_1\sqrt{l_s(l_s+2)}\rket{j_+;+}+\left[
g_0(l_s+2)-2g_1l_s\right]\rket{j_+;-} \ , 
\label{18b}
\eeq
\esub
\vspace{-0.6cm}
\bsub\label{19}
\beq
& &{\wtilde {\cal H}}\rket{j_-;-}=\left[
3g_0(l_s+1)-2g_1(l_s+1)\right]\rket{j_-;-}-2g_1(l_s+1)\rket{j_-;+} \ , 
\label{19a}\\
& &{\wtilde {\cal H}}\rket{j_-;+}=-2g_1(l_s+1)\rket{j_-;-}-\left[
g_0(l_s-1)+2g_1(l_s+1)\right]\rket{j_-;+} \ . 
\label{19b}
\eeq
\esub
Let $\rdket{j_+}$ and $\rdket{j_-}$ be the eigenstates of ${\wtilde {\cal H}}$: 
\beq
& &{\wtilde {\cal H}}\rdket{j_+}={\cal E}_{j_+}\rdket{j_+}\ , \qquad
\rdket{j_+}=\psi_+\rket{j_+;+}+\psi_-\rket{j_+;-}\ , 
\label{20}\\
& &{\wtilde {\cal H}}\rdket{j_-}={\cal E}_{j_-}\rdket{j_-}\ , \qquad
\rdket{j_-}=\phi_-\rket{j_-;-}+\phi_+\rket{j_-;+}\ . 
\label{21}
\eeq
The relations (\ref{18})$-$(\ref{21}) give the linear equations for the eigenvalue equations: 
\bsub\label{22}
\beq
&{\rm (i)}&\ {\rm for}\ j_+\hbox{\rm -states ; }\nonumber\\
& &\qquad
-[3g_0l_s+2g_1(l_s+2)]\psi_+-2g_1\sqrt{l_s(l_s+2)}\psi_-={\cal E}_{j_+}\psi_+\ , 
\label{22a}\\
& &\qquad
-2g_1\sqrt{l_s(l_s+2)}\psi_++[g_0(l_s+2)-2g_1l_s]\psi_-={\cal E}_{j_+}\psi_- \ , 
\label{22b}
\eeq
\esub
\vspace{-0.6cm}
\bsub\label{23}
\beq
&{\rm (ii)}&\ {\rm for}\ j_-\hbox{\rm -states ; }\nonumber\\
& &\qquad
[3g_0(l_s+1)-2g_1(l_s+1)]\phi_- -2g_1(l_s+1)\phi_+={\cal E}_{j_-}\phi_-\ , 
\label{23a}\\
& &\qquad
-2g_1(l_s+1)\phi_- -[g_0(l_s-1)+2g_1(l_s+1)]\phi_+={\cal E}_{j_-}\phi_+ \ . 
\label{23b}
\eeq
\esub
The normalization is performed by 
\beq\label{24}
\psi_+^2+\psi_-^2=1\ , \qquad 
\phi_-^2+\phi_+^2=1\ . 
\eeq

Let us start by searching solutions of the eigenvalue equations (\ref{22}) and (\ref{23}). 
We specify two solutions by the indices 1 and 2. 
First, we consider the case $g_0=0$. 
In this case, we have the following solutions for the eigenvalue equation: 
\bsub\label{25}
\beq
&{\rm (1a)}&\ \ {\cal E}_{j_+}^{(1)}=-4g_1(l_s+1)\ , \nonumber\\
& &\qquad
\rdket{j_+;1}=\sqrt{\frac{l_s+2}{2(l_s+1)}}\rket{j_+;+}+\sqrt{\frac{l_s}{2(l_s+1)}}\rket{j_+;-}\nonumber\\
& &\qquad\qquad\quad
=\frac{1}{2\sqrt{3(l_s+1)}}\left({\wtilde B}_{l_sj_+m\tau}^*(+)+{\wtilde B}_{l_sj_+m\tau}^*(-)\right)\rket{0} \nonumber\\
& &\qquad\qquad\quad
=\frac{1}{2\sqrt{3(l_s+1)}}{\wtilde B}_{l_sj_+m\tau}^{*(A)}\rket{0}\ . 
\label{25a}
\eeq
Here, we used the relation (\ref{12a}). 
\beq
&{\rm (1b)}&\ \ {\cal E}_{j_+}^{(2)}=0\ , \nonumber\\
& &\qquad
\rdket{j_+;2}=\sqrt{\frac{l_s+2}{2(l_s+1)}}\rket{j_+;+}-\sqrt{\frac{l_s}{2(l_s+1)}}\rket{j_+;-}\nonumber\\
& &\qquad\qquad\quad
=\frac{1}{2\sqrt{3(l_s+1)}}\left({\wtilde B}_{l_sj_+m\tau}^*(+)-{\wtilde B}_{l_sj_+m\tau}^*(-)\right)\rket{0} \ . 
\label{25b}
\eeq
\esub
For the eigenvalue equation (\ref{23}), we have the following solutions: 
\bsub\label{26}
\beq
&{\rm (2a)}&\ \ {\cal E}_{j_-}^{(1)}=-4g_1(l_s+1)\ , \nonumber\\
& &\qquad
\rdket{j_-;1}=\frac{1}{\sqrt{2}}\rket{j_-;-}+\frac{1}{\sqrt{2}}\rket{j_-;+}\nonumber\\
& &\qquad\qquad\quad
=\frac{1}{2\sqrt{3(l_s+1)}}\left({\wtilde B}_{l_sj_-m\tau}^*(-)+{\wtilde B}_{l_sj_-m\tau}^*(+)\right)\rket{0} \nonumber\\
& &\qquad\qquad\quad
=\frac{1}{2\sqrt{3(l_s+1)}}{\wtilde B}_{l_sj_-m\tau}^{*(A)}\rket{0}\ . 
\label{26a}
\eeq
Here, we used the relation (\ref{12b}). 
\beq
&{\rm (2b)}&\ \ {\cal E}_{j_-}^{(2)}=0\ , \nonumber\\
& &\qquad
\rdket{j_-;2}=\frac{1}{\sqrt{2}}\rket{j_-;-}-\frac{1}{\sqrt{2}}\rket{j_-;+}\nonumber\\
& &\qquad\qquad\quad
=\frac{1}{2\sqrt{3(l_s+1)}}\left({\wtilde B}_{l_sj_-m\tau}^*(-)-{\wtilde B}_{l_sj_-m\tau}^*(+)\right)\rket{0} \ . 
\label{26b}
\eeq
\esub
We can see that the states $\rdket{j_{\pm};1}$ are lower than $\rdket{j_{\pm};2}$ 
in energy and the lower states $\rdket{j_{\pm};1}$ are expressed in the form 
${\wtilde B}_{l_sj_{\pm}m\tau}^{*(A)}\rket{0}$. 
Since the above is the case in which the spin-orbit interaction is switched off, 
the above solution is quite natural and the Hamiltonian (\ref{16}) reproduces the 
above feature. 
In addition, we must mention that in any case with $g_0\neq 0$, we cannot derive the forms 
(\ref{25a}) and (\ref{26a}).

Next, we investigate the case $g_1=0$. 
As is clear from the eigenvalue equations (\ref{22}) and (\ref{23}), in this case, 
the two levels specified by $+$ and $-$ do not couple with each other. 
The eigenvalue equation (\ref{22}) gives the following solution: 
\bsub\label{27}
\beq
&{\rm (3a)}&\ \ {\cal E}_{j_+}^{(1)}=-3g_0l_s\ , \nonumber\\
& &\qquad
\rdket{j_+;1}=\rket{j_+;+}=
\frac{1}{\sqrt{6(l_s+2)}}{\wtilde B}_{l_sj_+m\tau}^*(+)\rket{0}\nonumber\\
& &\qquad\qquad\quad
=\frac{1}{\sqrt{6(l_s+2)}}{\wtilde B}_{l_sj_+m\tau}^{*(B)}\rket{0} \ . 
\label{27a}
\eeq
Here, we used the relation (\ref{13a}). 
\beq
&{\rm (3b)}&\ \ {\cal E}_{j_+}^{(2)}=g_0(l_s+2)\ , \nonumber\\
& &\qquad
\rdket{j_+;2}=\rket{j_+;-}=
\frac{1}{\sqrt{6l_s}}{\wtilde B}_{l_sj_+m\tau}^*(-)\rket{0} \ . \qquad\quad
\label{27b}
\eeq
\esub
Since ${\cal E}_{j_+}^{(1)}<{\cal E}_{j_+}^{(2)}$, the state (\ref{27a}) is nothing but the state we 
are looking for. 
The eigenvalue equation (\ref{23}) leads to the following solution: 
\bsub\label{28}
\beq
&{\rm (4a)}&\ \ {\cal E}_{j_-}^{(1)}=3g_0(l_s+1)\ , \nonumber\\
& &\qquad
\rdket{j_-;1}=\rket{j_-;-}=
\frac{1}{\sqrt{6(l_s+1)}}{\wtilde B}_{l_sj_-m\tau}^*(-)\rket{0}\nonumber\\
& &\qquad\qquad\quad
=\frac{1}{\sqrt{6(l_s+1)}}{\wtilde B}_{l_sj_-m\tau}^{*(B)}\rket{0} \ . 
\label{28a}
\eeq
Here, we used the relation (\ref{13b}). 
\beq
&{\rm (4b)}&\ \ {\cal E}_{j_-}^{(2)}=-g_0(l_s-1)\ , \nonumber\\
& &\qquad
\rdket{j_-;2}=\rket{j_-;+}=
\frac{1}{\sqrt{6(l_s+1)}}{\wtilde B}_{l_sj_-m\tau}^*(+)\rket{0} \ . 
\label{28b}
\eeq
\esub
Different from the previous three cases, the present gives us the relation 
${\cal E}_{j_-}^{(1)}>{\cal E}_{j_-}^{(2)}$. 
Therefore, we have to conclude that certainly we have the state related to 
the form (\ref{13b}), but, we find a color-singlet state which is lower than the state 
we intend to search.

The above analysis suggests us the following: 
The Hamiltonian (\ref{16}) can produce the color-singlet three-quark states 
and in energy, one of them is lower than the state generated by the operators 
(\ref{12}) and (\ref{13}). 
In this sense, the Hamiltonian (\ref{16}) is unsatisfactory for our purpose. 
By adding the third part to ${\wtilde {\cal H}}$, we adopt the 
following Hamiltonian: 
\beq
& &{\wtilde H}={\wtilde {\cal H}}+{\wtilde H}_2\ , 
\label{29}\\
& &{\wtilde H}_2=-g_2\left({\wtilde N}(+)-{\wtilde N}(-)\right)
\left(\sum_{i=1}^3{\wtilde S}^i(l_s){\wtilde S}_i(l_s)\right)
\left({\wtilde N}(+)-{\wtilde N}(-)\right)\ . 
\label{30}
\eeq
Here, $g_2$ denotes a parameter. 
Judging from the interactions, it may be permitted to set up the conditions 
$g_0\geq 0$ and $g_1\geq 0$ beforehand. 
But, ${\wtilde H}_2$ has an effect on ${\wtilde H}_0$ and ${\wtilde H}_1$. 
Therefore, beforehand, it may be impossible to specify that $g_2$ is positive or negative. 
The operation of ${\wtilde H}$ on the states (\ref{14}) and (\ref{15}) 
gives us the following result: 
\bsub\label{31}
\beq
{\wtilde H}\rket{j_+;+}&=&
-\left[3g_0l_s+2(g_1+9g_2)(l_s+2)\right]\rket{j_+;+}
\nonumber\\
& &
-2(g_1-3g_2)\sqrt{l_s(l_s+2)}\rket{j_+;-} \ , 
\label{31a}\\
{\wtilde H}\rket{j_+;-}&=&
-2(g_1-3g_2)\sqrt{l_s(l_s+2)}\rket{j_+;+}\nonumber\\
& &+\left[g_0(l_s+2)-2(g_1+g_2)l_s\right]\rket{j_+;-} \ , \qquad\ 
\label{31b}
\eeq
\esub
\vspace{-0.6cm}
\bsub\label{32}
\beq
{\wtilde H}\rket{j_-;-}&=&
\left[3g_0-2(g_1+9g_2)\right](l_s+1)\rket{j_-;-}
\nonumber\\
& &
-2(g_1-3g_2)(l_s+1)\rket{j_-;+} \ , 
\label{32a}\\
{\wtilde H}\rket{j_-;+}&=&
-2(g_1-3g_2)(l_s+1)\rket{j_-;-}\nonumber\\
& &-\left[g_0(l_s-1)+2(g_1+g_2)(l_s+1)\right]\rket{j_-;+} \ . 
\label{32b}
\eeq
\esub
With the use of the relations (\ref{20}) and (\ref{21}), in which 
${\wtilde {\cal H}}$ and ${\cal E}_{j_{\pm}}$ are replaced with 
${\wtilde H}$ and $E_{j_\pm}$, we have the following eigenvalue equations: 
\bsub\label{33}
\beq
&{\rm (i)}'&\ {\rm for}\ j_+\hbox{\rm -states ;} \nonumber\\
& &\quad
-\left[3g_0l_s+2(g_1+9g_2)(l_s+2)\right]\psi_+
-2(g_1-3g_2)\sqrt{l_s(l_s+2)}\psi_-=E_{j_+}\psi_+, \qquad
\label{33a}\\
& &\quad
-2(g_1-3g_2)\sqrt{l_s(l_s+2)}\psi_+
+\left[g_0(l_s+2)-2(g_1+g_2)l_s\right]\psi_-
=E_{j_+}\psi_-\ , 
\label{33b}
\eeq
\esub
\vspace{-0.6cm}
\bsub\label{34}
\beq
&{\rm (ii)}'&\ {\rm for}\ j_-\hbox{\rm -states ;} \nonumber\\
& &\quad
\left[3g_0-2(g_1+9g_2)\right](l_s+1)\phi_-
-2(g_1-3g_2)(l_s+1)\phi_+=E_{j_-}\phi_-\ , 
\label{34a}\\
& &\quad
-2(g_1-3g_2)(l_s+1)\phi_-
-\left[g_0(l_s-1)+2(g_1+g_2)(l_s+1)\right]\phi_+
=E_{j_-}\phi_+\ . \qquad\ \ 
\label{34b}
\eeq
\esub
Of course, if $g_2=0$, the relations (\ref{33}) and (\ref{34}) are reduced 
to the relations (\ref{22}) and (\ref{23}), respectively and the condition 
(\ref{24}) is also used in the present case.

Let us investigate the eigenvalue equations (\ref{33}) and (\ref{34}). 
First, we search the condition which leads to the same solution as that 
shown by the relation (\ref{25a}). 
By substituting $\psi_+=\sqrt{(l_s+2)/2(l_s+1)}$ 
and 
$\psi_-=\sqrt{l_s/2(l_s+1)}$ into Eq.(\ref{33}), we have the following condition: 
\beq\label{35}
g_0(2l_s+1)+8\left(l_s+3\right)g_2=0 \ . 
\eeq
Further, by substituting $\phi_+=1/\sqrt{2}$ and $\phi_-=1/\sqrt{2}$ into Eq.(\ref{34}), 
we have 
\beq\label{36}
g_0(2l_s+1)-12g_2(l_s+1)=0 \ . 
\eeq
Therefore, the common solution to the conditions (\ref{35}) and (\ref{36}) is given 
by 
\beq\label{37}
g_0=0\ , \qquad g_2=0\ .
\eeq
The solution (\ref{37}) gives the results which are the same as those 
presented in (1a), (1b), (2a) and (2b) in Eqs.(\ref{25}) and (\ref{26}). 
For the case $g_0=g_2=0$, we can summarize the result as follows: 
\bsub\label{38}
\beq
& &{\rm (1a)}'\ : \ {\rm (1a)}\ , 
\label{38a}\\
& &{\rm (1b)}'\ : \ {\rm (1b)}\ , 
\label{38b}
\eeq
\esub
\vspace{-0.6cm}
\bsub\label{39}
\beq
& &{\rm (2a)}'\ : \ {\rm (2a)}\ , 
\label{39a}\\
& &{\rm (2b)}'\ : \ {\rm (2b)}\ . 
\label{39b}
\eeq
\esub

Next, we investigate the case where 
the spin-orbit partner levels are uncoupled with each other. 
As is clear from the eigenvalue equation (\ref{33}), 
two levels become uncoupled with each other under the condition 
\beq\label{40}
g_1-3g_2=0\ , \quad {\rm i.e.,}\quad 
g_1=3g_2\ . \quad (g_2 \geq 0)
\eeq
In this case, we have 
\bsub\label{41}
\beq
& &E_{j_+}^{(1)}=-3g_0l_s-24g_2(l_s+2)\ , 
\label{41a}\\
& &E_{j_+}^{(2)}=g_0(l_s+2)-8g_2l_s\ . 
\label{41b}
\eeq
\esub
Since $E_{j_+}^{(2)}-E_{j_+}^{(1)}=2g_0(2l_s+1)+16g_2(l_s+3)>0$, the state with 
$E_{j_+}^{(1)}$ is nothing but the state we are looking for. 
For the states with $j_-$, the relation (\ref{40}) is also valid. 
In this case, the eigenvalue equation (\ref{34}) gives us 
\bsub\label{42}
\beq
& &E_{j_-}^{(1)}=(3g_0-24g_2)(l_s+1)\ , 
\label{42a}\\
& &E_{j_-}^{(2)}=-g_0(l_s-1)-8g_2(l_s+1)\ . 
\label{42b}
\eeq
\esub
Therefore, we have 
\beq\label{43}
E_{j_-}^{(2)}-E_{j_-}^{(1)}=-2\left[
g_0(2l_s+1)-8g_2(l_s+1)\right] \ .
\eeq
Since the state, which we are looking for, should obey the condition 
$E_{j_-}^{(1)}<E_{j_-}^{(2)}$, the relation (\ref{43}) gives us 
\beq\label{44}
g_0<8g_2\cdot \frac{l_s+1}{2l_s+1}\ .
\eeq
It should be noted that the inequality $E_{j_-}^{(1)}<E_{j_-}^{(2)}$ cannot be 
derived under the Hamiltonian (\ref{16}). 
Thus, for the case with $g_1=3g_2$ and $g_0<8g_2\cdot (l_s+1)/(2l_s+1)$, the results are summarized 
as follows: 
\bsub\label{45}
\beq
&{\rm (3a)}'&\ E_{j_+}^{(1)}=-3g_0l_s-24g_2(l_s+2)\ , \nonumber\\
& &\quad
\rdket{j_+;1}=\frac{1}{\sqrt{6(l_s+2)}}{\wtilde B}_{l_sj_+m\tau}^*(+)\rket{0}
=\frac{1}{\sqrt{6(l_s+2)}}{\wtilde B}_{l_sj_+m\tau}^{*(B)}\rket{0} \ , 
\label{45a}\\
&{\rm (3b)}'&\ E_{j_+}^{(2)}=g_0(l_s+2)-8g_2l_s\ , \nonumber\\
& &\quad
\rdket{j_+;2}=\frac{1}{\sqrt{6l_s}}{\wtilde B}_{l_sj_+m\tau}^*(-)\rket{0} \ , 
\label{45b}
\eeq
\esub
\vspace{-0.6cm}
\bsub\label{46}
\beq
&{\rm (4a)}'&\ E_{j_-}^{(1)}=3g_0(l_s+1)-24g_2(l_s+1)\ , \nonumber\\
& &\quad
\rdket{j_-;1}=\frac{1}{\sqrt{6(l_s+1)}}{\wtilde B}_{l_sj_-m\tau}^*(-)\rket{0}
=\frac{1}{\sqrt{6(l_s+1)}}{\wtilde B}_{l_sj_-m\tau}^{*(B)}\rket{0} \ , 
\label{46a}\\
&{\rm (4b)}'&\ E_{j_-}^{(2)}=-g_0(l_s-1)-8g_2(l_s+1)\ , \nonumber\\
& &\quad
\rdket{j_-;2}=\frac{1}{\sqrt{6(l_s+1)}}{\wtilde B}_{l_sj_-m\tau}^*(+)\rket{0} \ . 
\label{46b}
\eeq
\esub
As can be seen in the relation (\ref{46}), the Hamiltonian (\ref{29}) solved the 
problem which was brought about by the Hamiltonian (\ref{16}).

It may be also important to consider the color-singlet three-quark states 
in the $\Delta$-excitation. 
First, the state described by the second quark-triplet operator in (I$\cdot$6$\cdot$6) is given by 
\beq\label{47}
{\wtilde D}_{l_s\mu\sigma\tau}^*\rket{0}=\prod_{i=1}^3{\tilde c}_{il_s\mu\sigma\tau}^*\rket{0}\ . 
\eeq
In the case $g_0=g_2=0$, we have 
\beq\label{48}
{\wtilde H}{\wtilde D}_{l_s\mu\sigma\tau}^*\rket{0}=E_{l_s}^{(\Delta)}{\wtilde D}_{l_s\mu\sigma\tau}^*\rket{0} \ , 
\qquad E_{l_s}^{(\Delta)}=0\ .
\eeq
The state ${\wtilde D}_{l_s\mu\sigma\tau}^*\rket{0}$ is higher than 
$\rdket{j_+;1}$ and $\rdket{j_-;1}$ shown by relations (\ref{38a}) and (\ref{39a}). 
Next, we investigate the case described by the second quark-triplet operator in (I$\cdot$6$\cdot$3):
\beq\label{49}
{\wtilde D}_{l_sj_{\pm}m\tau}^*\rket{0}=\prod_{i=1}^3{\tilde c}_{il_sj_{\pm}m\tau}^*\rket{0} \ .
\eeq
This case gives  
\bsub\label{50}
\beq
& &{\wtilde H}{\wtilde D}_{l_sj_+m\tau}^*\rket{0}
=E_{j_+}^{(\Delta)}{\wtilde D}_{l_sj_+m\tau}^*\rket{0}\ , \qquad
E_{j_+}^{(\Delta)}=-3g_0l_s\ , 
\label{50a}\\
& &{\wtilde H}{\wtilde D}_{l_sj_-m\tau}^*\rket{0}
=E_{j_-}^{(\Delta)}{\wtilde D}_{l_sj_-m\tau}^*\rket{0}\ , \qquad
E_{j_-}^{(\Delta)}=3g_0(l_s+1)\ . 
\label{50b}
\eeq
\esub
We require that ${\wtilde D}_{l_sj_{\pm}m\tau}^*\rket{0}$ is higher than 
$\rdket{j_+;1}$ and $\rdket{j_-;1}$ shown by the relation (\ref{45a}) and (\ref{46a}), 
respectively: 
\bsub\label{51}
\beq
& &E_{j_+}^{(\Delta)}-E_{j_+}^{(1)}=24g_2(l_s+2)>0\ , 
\label{51a}\\
& &E_{j_+}^{(\Delta)}-E_{j_-}^{(1)}=-3\left[g_0(2l_s+1)-8g_2(l_s+1)\right] >0\ , 
\label{51b}
\eeq
\esub
\vspace{-0.6cm}
\bsub\label{52}
\beq
& &E_{j_-}^{(\Delta)}-E_{j_+}^{(1)}=3\left[g_0(2l_s+1)+8g_2(l_s+1)\right] >0\ , \ 
\label{52a}\\
& &E_{j_-}^{(\Delta)}-E_{j_-}^{(1)}=24g_2(l_s+1)>0\ . 
\label{52b}
\eeq
\esub
Since $g_2>0$, the inequalities (\ref{51a}), (\ref{52a}) and (\ref{52b}) are automatically 
satisfied. 
Under the condition (\ref{44}), we have the inequality (\ref{51b}). 
The above is the comparison with the $\Delta$-excitation.

The arguments presented in this note lead to the following conclusion 
for one ``nucleon" states: 
\bsub\label{53}
\beq
{\rm (A)}& &\ {\rm The\ case}\ (g_0=0,\ g_2=0) ; \nonumber\\
&(1)&\ \ \rdket{j_+;1}=\frac{1}{2\sqrt{3(l_s+1)}}
{\wtilde B}_{l_sj_+m\tau}^{*(A)}\rket{0}\ , 
\nonumber\\
& &\ \ E_{j_+}^{(1)}=-4g_1(l_s+1)\ , 
\label{53a}\\
&(2)&\ \ \rdket{j_-;1}=\frac{1}{2\sqrt{3(l_s+1)}}
{\wtilde B}_{l_sj_-m\tau}^{*(A)}\rket{0}\ , 
\nonumber\\
& &\ \ E_{j_-}^{(1)}=-4g_1(l_s+1)\ , 
\qquad\qquad\qquad\qquad\qquad\qquad\qquad\qquad
\label{53b}
\eeq
\esub
\vspace{-0.6cm}
\bsub\label{54}
\beq
{\rm (B)}& &\ {\rm The\ case}\ (g_0<8g_2(l_s+1)/(2l_s+1),\ g_1=3g_2) ; \nonumber\\
&(3)&\ \ \rdket{j_+;1}=\frac{1}{\sqrt{6(l_s+2)}}{\wtilde B}_{l_sj_+m\tau}^{*(B)}\rket{0}\ , 
\nonumber\\
& &\ \ E_{j_+}^{(1)}=-3g_0l_s-24g_2(l_s+2)\ , \qquad\qquad\qquad\qquad\qquad\qquad
\label{54a}\\
&(4)&\ \ \rdket{j_-;1}=\frac{1}{\sqrt{6(l_s+1)}}{\wtilde B}_{l_sj_-m\tau}^{*(B)}\rket{0}\ , 
\nonumber\\
& &\ \ E_{j_-}^{(1)}=3g_0(l_s+1)-24g_2(l_s+1)\ . 
\label{54b}
\eeq
\esub
The energies $E_{j_+}^{(1)}$ and $E_{j_-}^{(1)}$ given by the relation (\ref{54}) 
can be rewritten in the form 
\beq
& &E_{j_+}^{(1)}=E_{l_0}-\xi_{l_0}l_s\ , \qquad
E_{j_-}^{(1)}=E_{l_0}+\xi_{l_0}(l_s+1)\ , 
\label{55}\\
& &E_{l_0}=-48g_2\frac{(l_s+1)^2}{2l_s+1}\ , \qquad
\xi_{l_0}=3g_0+24g_2\frac{1}{2l_s+1}\ . 
\label{56}
\eeq
The energy $E_{l_0}$ tells that the origin measuring the energy changes from 
the point 0, which we adopted at the starting point. 
Of course, this change arises as a consequence of ${\wtilde H}_2$. 
The symbol $\xi_{l_0}$ indicates the strength of the spin-orbit 
interaction for one ``nucleon" state. 
It may be interesting to see that the value deviates from 
$3g_0$, the value of three bare quarks. 
It also comes from the effect of ${\wtilde H}_2$. 

Thus, we have the following conclusion: 
If the value of the parameters contained in the Hamiltonian (\ref{29}) are 
chosen appropriately, we can derive the color-singlet three-quark states 
presented in Ref.\citen{2} for both of the spin-orbit partner levels. 
Therefore, it may be interesting to investigate the deviation from the 
states (\ref{54a}) and (\ref{54b}). 
It is our next task. 
In Ref.\citen{3}, the states (\ref{54a}) and (\ref{54b}) were applied to the 
problem of nuclear magnetic moments and various discussions were performed. 
Since the present note does not aim at the application, for example, 
we does not contact with the problems 
related to those discussed in Ref.\citen{3}.

\vspace{1cm}

One of the authors (Y.T.) 
is partially supported by the Grants-in-Aid of the Scientific Research 
(No.23540311) from the Ministry of Education, Culture, Sports, Science and 
Technology in Japan.


\end{document}